\documentclass[aps,prl,showpacs,twocolumn,citeautoscript,superscriptaddress]{revtex4-1}

\usepackage{graphics}
\usepackage{bigints}
\usepackage{graphicx}
\usepackage{mathrsfs}
\usepackage{amsmath,amssymb}
\usepackage{bm}
\usepackage{color}
\usepackage{float}
\usepackage{comment}

\begin{document}
	
\title{A Unified Model for Non-Fickian Diffusion and Anomalous Swelling of Glassy Polymer Gels}
\author{Peihan Lyu}
\affiliation{School of Physics, Beihang University, Beijing 100191, China}
\author{Zhaoyu Ding}
\affiliation{School of Physics, Beihang University, Beijing 100191, China}
\author{Masao Doi}
\email{masao.doi@buaa.edu.cn}
\affiliation{School of Physics, Beihang University, Beijing 100191, China}
\affiliation{Wenzhou Institute, University of Chinese Academy of Sciences, Wenzhou 325000, China}
\author{Xingkun Man}
\email{manxk@buaa.edu.cn}
\affiliation{School of Physics, Beihang University, Beijing 100191, China}
\affiliation{Peng Huanwu Collaborative Center for Research and Education, Beihang University, Beijing 100191, China}

\begin{abstract}
A sheet of glassy polymers placed in solvent shows swelling behaviors 
quite different from that of soft polymers (rubbers and gels).
(1) Non-Fickian diffusion (called case II diffusion): As solvent permeates into
the sample,  a sharp front is created between the swollen part 
and the glassy part, and it moves towards the center at constant speed. 
(2) Non-monotonous swelling: The thickness of 
the sample first increases and then decreases towards the equilibrium value.   
Here we propose a theory to explain such anomalous behavior by extending the previous theory for swelling of soft gels. We regard the material as a continuum mixture of glassy polymer network and solvent. We assume that the polymer network is a viscoelastic gel of glassy polymers and its relaxation time depends strongly on solvent concentration.
We will show that this theory explains the above two characteristics of glassy polymers
in a simple and unified framework. The theory predicts how the permeation 
speed of solvent and the characteristic times of the swelling process depend on material parameters and experimental conditions, which can be checked experimentally.

\end{abstract}
	
\maketitle
	
When a thin sample of polymer is placed in a bath of solvent, 
the solvent permeates into the polymer and increases the volume of the sample.  In soft polymers (rubbers and gels), the thickness and the area of 
the sample increase monotonically in time $t$, following the Fickian 
diffusion law (in proportion  to $t^{1/2}$) and reach the equilibrium value.
On the other hand, in glassy polymer, the swelling behavior is
completely different~\cite{TW1977,alfrey1966First,Sanopoulou2001nonFick,Ji2002CaseIIExp,DeKee2005NonFickianReview,PETROPOULOS2011nonFickModel}.  In the classical experiment,
Thomas and Windle \cite{TW1977}   
reported the following anomalies in the swelling of PMMA(poly-methyl-methaclylate) sheet in methanol (see Fig.1).
(a) At first, as the solvent permeates into the polymer, a soft rubbery 
 part  is created at the surface of the polymer and moves towards the center.  During this process, the interface between the rubbery part and the glassy part remains sharp, and it moves at constant speed.  At this stage, 
 the sample thickness increases in time in proportion to $t$, while the sample width remains constant. (b) After some time, the two fronts created at both surfaces meet at the center and the whole polymer becomes rubbery.  At this stage,  the thickness decreases in time while the width increases towards their equilibrium values. 

The non-Fickian diffusion dynamics characterizing the first stage is 
observed for many glassy polymers, and is called case II diffusion~\cite{TW1981DiffusionMO,LASKY1988Temperature,Foreman2015AnomalousExp,Nixdorf2019CaseIIExp,Bischak2020,Hausmann2023CaseII}. The case II diffusion 
is characterized by the existence of sharp front  of swollen region that 
moves at constant speed. Such behavior is considered to be caused by the 
glass transition of polymers induced by solvent.  Numerous theories have
been proposed to describe the case II  diffusion~\cite{TW1980,Rossi1995CaseII,QIAN2000CaseII,Bargmann2011review,Miao2017JCP,Borrmann2021iecr,SONG2023CaseIIModelFEM}. 
However, there is still no consensus or clear understanding on how 
the case II behavior is caused by the glass transition.  

All previous theories have a common drawback that they only discuss 
the one dimensional problem  of diffusion in the first stage~\cite{TW1982CaseII,Kalospiros1991GeneralNonFick,
Jou1991EIT,LUSTIG1992model,Govindjee1993CaseII,WILMERS2015CaseII,Krenn2020MaxwellStefan,crank1975book,Carbonell1990SlabSwell}.  Such theory 
cannot discuss the three dimensional problem of shape change in the second stage and how the first stage transforms to the second stage.
To deal with the second stage, separate theory was needed.

%
\begin{figure}[bp]
	\begin{center}
	\includegraphics[width=0.5\textwidth,draft=false]{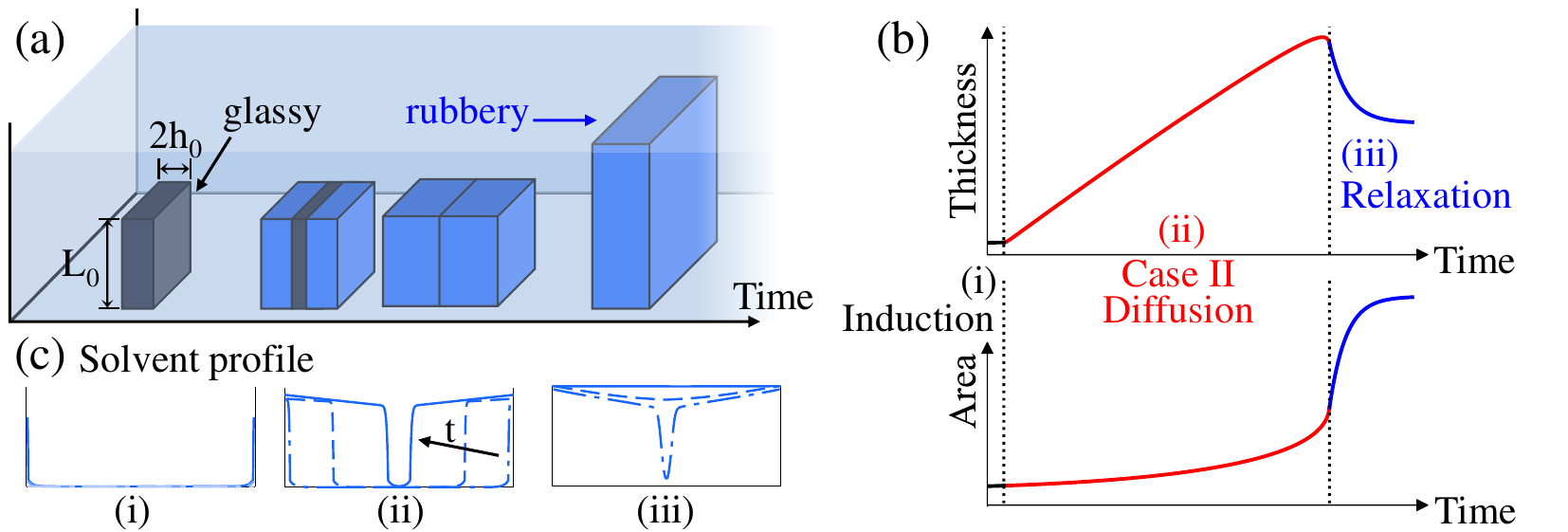}
	\caption{(a) Schematic of the swelling of a square sheet of glassy polymers in a bath of solvent. The sheet initially has side length $L_0$ and thickness $2h_0$ ($h_0 \ll L_0$). As solvent permeates, sharp front is formed between outer rubbery part (blue) and inner glassy part (black). After some time, the entire sample becomes rubbery, relaxing towards the equilibrium state. (b) Time evolution of thickness (top) and area (bottom) show three stages. (i) Induction stage (black): both thickness and area are nearly unchanged. (ii) Case II diffusion stage (red): thickness increases in time in proportion to $t$, while area nearly remains constant. (iii) Relaxation stage (blue): thickness decreases but area increases towards the equilibrium value. (c) Time evolution of the solvent profile along $z$-axis in the three stages.
	}
	\label{fig1}
	\end{center}
\end{figure}

In this paper,  we propose a three dimensional theory 
for the swelling of glassy polymer. We regard the glassy polymer as a viscoelastic gel  of glassy polymer and discuss its swelling using the same framework as
that for the soft polymer gels~\cite{doi2009gelreview, man2021swelling, ding2022diffusio,lyu2023swelling}. 
We take the conventional view for glass transition, 
and  assume that the kinetic parameters (solvent diffusion constants and polymer relaxation times) change by orders of magnitude at some characteristic solvent volume fraction.
We shall show that this  simple model reproduces  the two anomalies, 
the case II diffusion and the non-monotonous shape change without introducing
any additional assumptions.

We consider a square sheet of glassy polymer gel that is initially homogeneous and isotropic
with thickness $2h_0$ and side length $L_0$ ($h_0 \ll L_0$), (see  Fig.1). We take a 
Cartesian coordinate such that the right and the left surfaces 
are at $z = \pm h_0$. Suppose that a point located at $\bm{r}$ on 
the polymer network in the dry state is displaced 
to $\bm{r}+  \bm {u}(\bm{r},t)$ at time $t$. We use a homogeneous continuum model and 
assume that the displacement gradient $\varepsilon_{\alpha \beta} = \partial u_{\alpha}/\partial r_{\beta}$ 
($\alpha, \beta$ stand for $x,y,z$) is small, and that the swelling takes place uniformly and isotropically in $x$-$y$ plane. The displacement vector can be written as
$u_x(t) = \alpha(t)x$, $u_y(t)= \alpha(t)y$, and  $u_z=u(z,t)$, where
$\alpha(t)$ and $u(z,t)$ are the unknowns to be obtained.  

For small deformation, the  polymer velocity $\bm{v}_p$ at $\bm{r}$ is given by the time derivative of $\bm {u} (\bm{r},t)$, i.e. $\bm{v}_p = \dot {\bm{u}}$. The solvent velocity $\bm{v}_s$ is different from $\bm{v}_p$ and the difference represents the diffusion (or permeation) of solvent in polymer network. We assume that the system is incompressible as
a whole, and always take the dry state as the reference state. This allows us to write the volume fraction of solvent $\phi_s$ in terms of the volume expansion of polymer network as $\phi_s=\nabla \cdot \bm{u} = \partial u/\partial z + 2 \alpha$. 
Therefore, the state of the sample at time $t$ is completely characterized by $u(z,t)$ and $\alpha(t)$.

To determine the time evolution of $u(z,t)$ and $\alpha(t)$, we use Onsager principle, which states that their evolution rate $\dot{u}$ and $\dot{\alpha}$ are determined by the minimum of the Rayleighian function defined by $\Re = \dot F + \Phi$. Here, $\dot F$ is the time derivative of free energy $F$, and $\Phi$ is the energy dissipation function of the system~\cite{doi2013soft, DOI2021OVP}. 

The free energy $F$ includes elastic energy of polymer network and mixing energy of polymer and solvent. For small deformation, the free energy is 
given in the same form as in soft gels~\cite{man2021swelling, ding2022diffusio} 
\begin{equation}\label{eqn:freeE}
  \dfrac{F}{L^2_0} = \int^{h_0}_0 {\rm d}z \left[\frac{K}{2}\left(2\alpha+\frac{\partial u}{\partial z}-3\varepsilon_{\rm eq} \right)^2+\frac{2G}{3}\left(\frac{\partial u}{\partial z}-\alpha\right)^2\right]
\end{equation}
where $\varepsilon_{\rm eq}$ is the equilibrium strain, and $K$ and $G$ are the material constants called osmotic bulk modulus and shear modulus, respectively.

The energy dissipation function has two parts. (a) $\Phi_{\rm dif}$ which arises from the friction between polymer and solvent. This term is given by the same as that for soft gels  $\Phi_{\rm dif} =\int {\rm d} \bm{r}  \frac{\xi}{2} (\bm{v}_p - \bm{v}_s)^2$,
where $\xi$ is the friction coefficient. (b) $\Phi_{\rm rhe}$ which arises from the
configurational change of the polymer. This is a new term introduced in the present
theory to account for the glass transition. In polymeric materials, the glass transition
is a rheological transition and has been represented by a very
large change of rheological relaxation  time across the transition point~\cite{Ferry}. To account for this effect, we model the polymer network of glassy gel by
 the Kelvin-Voigt model for viscoelastic solid. This model is derived from the energy dissipation function
$\Phi_{\rm rhe} =  \int {\rm d} \bm{r}  
	 \frac{\eta}{4} \sum_{\alpha \beta} 
	\left (\dot \varepsilon_{\alpha \beta} + \dot \varepsilon_{\beta \alpha } 
	- \frac{2}{3}
	\delta_{\alpha \beta} \sum_{\gamma} \dot\varepsilon_{\gamma \gamma}
	\right )^2$~\cite{SM}. 
Here $\eta$ is a material parameter called internal viscosity,  which is defined by the rheological relaxation time $\tau_{\rm rhe}$ as $\eta=G \tau_{\rm rhe}$.

Our model includes two kinetic parameters  $\xi$ and $\eta$ :
$\xi$  represents the dissipation due to the relative motion
between polymer and solvent, and $\xi$ represents the dissipation due to the
conformational change of the polymer. We assume that they  
change by orders of magnitude at some  solvent volume fraction 
$\phi_{\rm cri}$. Specifically, we assume 
that $\xi$ and $\eta$ depend on $\phi_s$ as $ \xi (\phi _s) = \xi _r + \frac{1}{2}\left( \xi _g - \xi _r \right)\left( 1 + \tanh \frac{\phi_{\rm cri} - {\phi_s}}{\phi _w} \right) $ and $\eta (\phi _s) = \eta _r + \frac{1}{2}\left( \eta _g - \eta _r \right)\left( 1 + \tanh \frac{\phi_{\rm cri} - {\phi_s}}{\phi _w} \right)$, where $\xi_g$ and $\eta_g$ ($\xi_r$ and $\eta_r$) are the friction coefficient and viscosity in the glassy state (rubbery state), and $\phi_w$ is the width of the glass transition region.  On the other
hand, we assume that the modulus 
$K$ and $G$ are constants independent of $\phi_s$.

%

\begin{figure}[tp]
	\begin{center}
	\includegraphics[width=0.48\textwidth,draft=false]{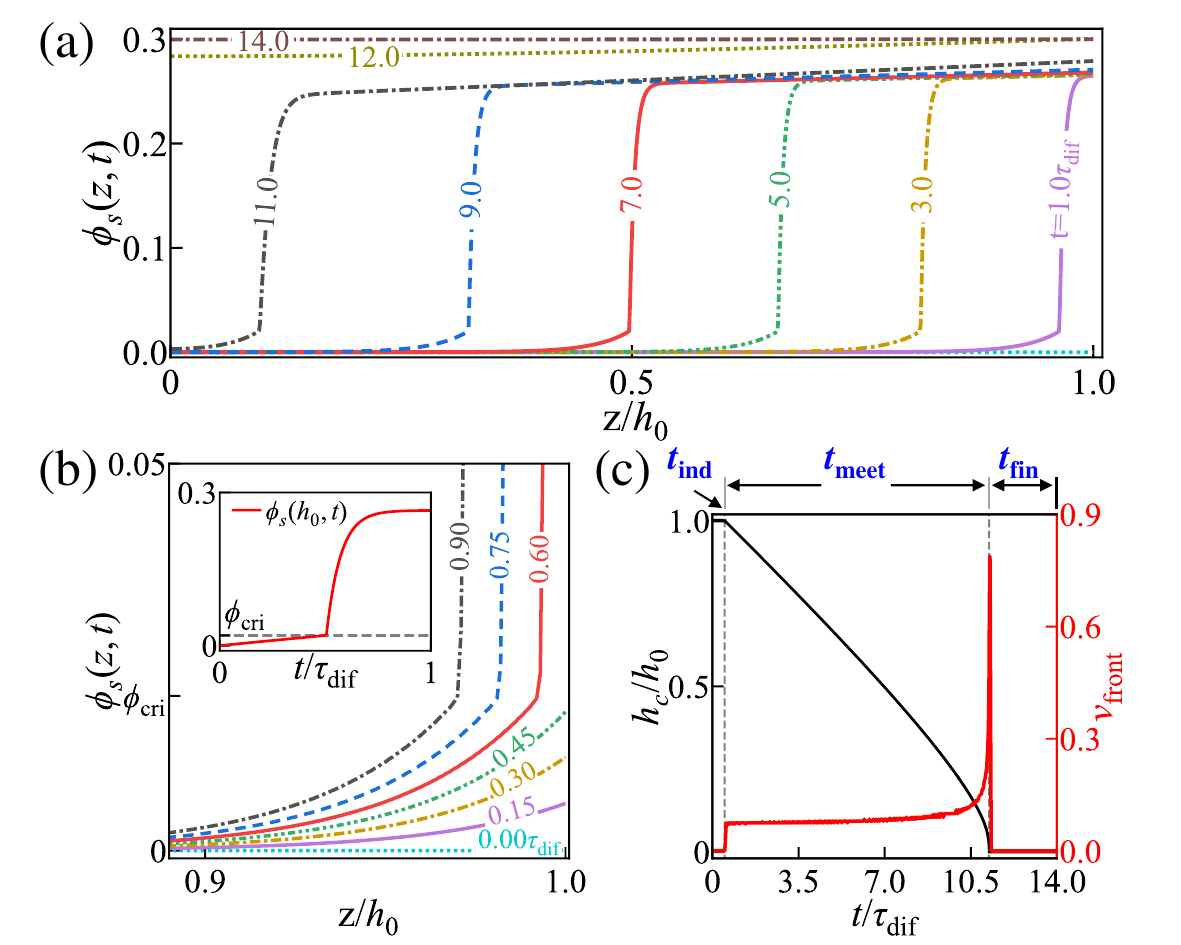}
	\caption{ 
	(a) Time evolution of the solvent volume fraction $\phi_s (z,t)$ in the swelling of a sample which includes no solvent at time $t=0$, i.e., $\phi(z,0)=0$. (b) Zoomed-in view of (a) in the region near the surface at early time of permeation. The inset is the time dependence of solvent volume fraction at gel surface $\phi_s(h_0, t)$. (c) Time dependence of the front position $h_c(t)$ and the front velocity $v_{\rm front}$, indicating three characteristic periods: (i) induction period, $t_{\rm ind}$, during which $h_c(t)$ remains unchanged at the surface, (ii) case II diffusion period, 
	$t_{\rm meet}$, during which $h_c$ moves towards the center at a nearly constant speed, and (iii) relaxation period $t_{\rm fin}$, during which $h_c$ decreases to zero. All times are scaled by $\tau_{\rm dif}=h_0^2/D_r$. Other parameters are $\xi_g/\xi_r=5000$, $\eta_g/\eta_r=100$ and $\eta_r/\xi_r h^2_0=0.05$.
	 }
	\label{fig2}
	\end{center}
\end{figure}

For the present problem, the energy dissipation function $\Phi = \Phi_{\rm dif} + \Phi_{\rm rhe}$ is written as \cite{SM}
 \begin{equation}\label{eqn:DissipationFunc}
 	\frac{\Phi}{L_0^2} = \int_{0}^{h_0} {\rm d}z \left[ \frac{\xi(\phi_s)}{2}\left(\dot u+2\dot \alpha z\right)^2 + \frac{2\eta(\phi_s)}{3}\left( \frac{\partial \dot u}{\partial z} - \dot\alpha \right)^2 \right]
 \end{equation} 
 
The Rayleighian of the system is given by 
$\Re=\Phi + \dot F$. The minimum condition of the Rayleighian gives  
the following set of equations\cite{SM}. Time evolution equation,
\begin{align}
	&\xi(\dot u+2\dot\alpha z) = ( K + \frac{4}{3}G )\frac{\partial^2 u}{\partial z^2} + \frac{4}{3}\frac{\partial }{\partial z}\left[ \eta\left( \frac{\partial \dot u}{\partial z}  - \dot \alpha  \right) \right]                      \label{eqn:EvoEqns1}\\
	& G \left( u - \alpha h_0 \right) + \int_0^{h_0} { \eta \left( \frac{\partial \dot u}{\partial z}  - \dot \alpha  \right){\rm{d}}z}  = 0   \label{eqn:EvoEqns2}       
\end{align}
and the boundary condition at $z=h_0$
\begin{equation}\label{eqn:BoundaryConds}
	\frac{4}{3}\eta \left( {\dot \phi_s - 3\dot \alpha } \right) + K\left( {\phi_s- 3{\varepsilon _{\rm eq}}} \right) + \frac{4}{3}G\left( {\phi_s - 3\alpha } \right) = 0
\end{equation}
Equations~\eqref{eqn:EvoEqns1} and~\eqref{eqn:EvoEqns2} are the set of equations which describe the swelling process of glassy polymers. 

Equation~\eqref{eqn:EvoEqns1} indicates that the motion in $z$ direction is determined by two terms,  the diffusion term $(K+\frac{4}{3}G)\frac{\partial \phi_s}{\partial z}$ and the viscosity term $\frac{4}{3}\frac{\partial}{\partial z}\left[\eta(\dot\phi_s-3\dot \alpha)\right]$. The diffusion pushes solvent towards glassy region, but the internal viscosity hinders such permeation.

The boundary condition (\ref{eqn:BoundaryConds}) is automatically derived from 
the variational calculus.  It represents the condition that the normal stress 
acting on the polymer network at $z=h_0$ is zero. This boundary condition is quite
different from the usual Neuman condition for soft 
gels~\cite{man2021swelling, ding2022diffusio}, and plays an
important role in the subsequent discussions.  Due to the symmetry of the system, 
we use the boundary condition $\partial u(z, t)/\partial z=0$ at $z=0$;

We solved eqs.~(\ref{eqn:EvoEqns1})-(\ref{eqn:BoundaryConds}) numerically, and calculated the time variation of the solvent volume fraction  
$\phi_s(z,t)= \partial u/\partial z + 2 \alpha$, the thickness change
$\Delta h(t)= u(h_0,t)$  and the area change $2 L_0^2\alpha(t)$. 

We first study the system that initially there is no solvent in sample, i.e. the initial value of the volume fraction of solvent $\phi_0$ is equal to $0$. 

Figure~2(a) shows the time evolution of the solvent volume fraction $\phi_s(z,t)$ obtained
by the numerical calculation.  Here $\tau_{\rm dif}=h_0^2/D_r$  with $D_r=(K+\frac{4}{3}G)/\xi_r$, represents the characteristic swelling time
~\cite{Tanaka1979swelling} of the gel in rubbery region and is taken to 
be the unit of time in this paper. It is seen that the solvent permeation proceeds with
characteristic step-function shape: the glassy region where 
$ \phi_s(z,t)$ is less than $\phi_{\rm cri}$ and the rubbery region where
where $ \phi_s(z,t)$ is much larger than $\phi_{\rm cri}$.
The interface between the two regions is sharp.  To see the motion 
of the interface, we define the front position $h_c(t)$ as the point where $\phi_s(z,t)$ is equal to 
$\phi_{\rm cri}$, and plotted $h_c(t)$ and its time derivative, i.e., the front velocity $v_{\rm front}
=\dot{h}_c(t)$, as a function of time in Figure 2(c). 

In the beginning,  there is no rubbery region and we define the front at the surface, i.e., 
$h_c(t)=h_0$.  At the late stage of swelling, the glassy region disappears, and we define 
$h_c(t)=0$.  In the following, we call the first 
period in which there is no rubbery region induction period, and denote its duration time with $t_{\rm ind}$. 
We call the last stage in which there is no glassy region, the relaxation period, and
denote its duration time by $t_{\rm fin}$.  We shall call the period between these two stages
as the case II period, and denote its duration time by $t_{\rm meet}$. We now discuss each period in more detail.

%
\begin{figure}[tp]
	\begin{center}
	\includegraphics[width=0.48\textwidth,draft=false]{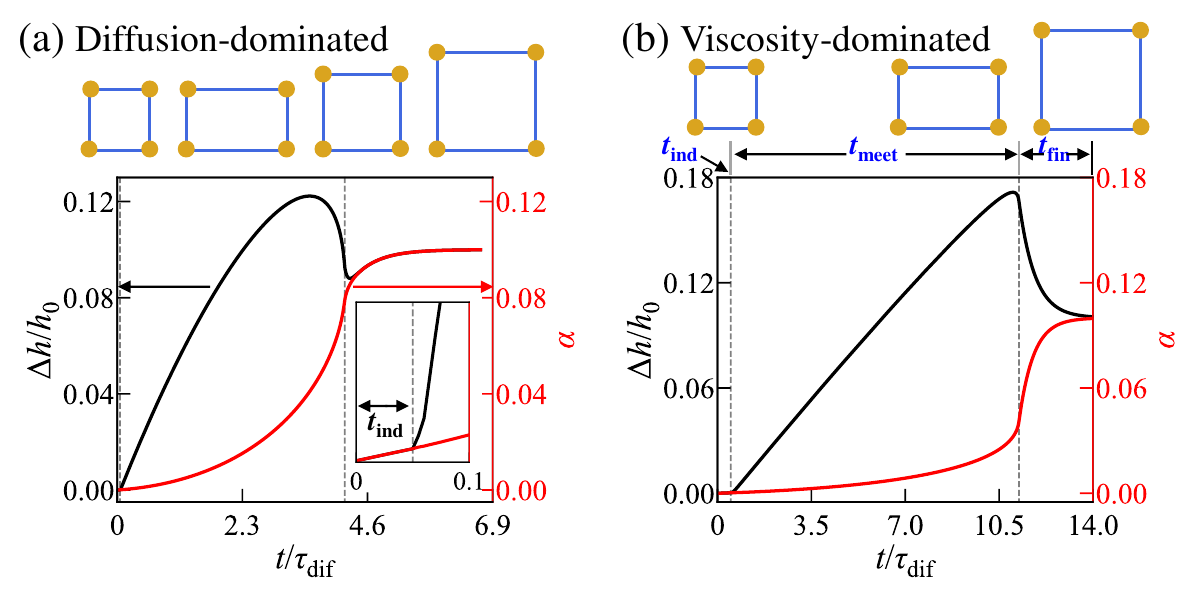}
	\caption{
		Time dependence of the two strains $\Delta h(t)/h_0$ (black solid line) and $\alpha(t)$ (red solid line) for glassy polymers with (a) lower viscosity ($\eta_r/\xi_r h_0^2=0.005$) and (b) higher viscosity ($\eta_r/\xi_r h_0^2=0.05$). The top figure shows schematic time evolution of the apparent shape of the sample. The last stage in (a) is diffusion-dominated and in (b) is viscosity-dominated. Other parameters are the same as in Fig.~2.
	 }
	\label{fig3}
	\end{center}
\end{figure}
%

{\bf Induction period.} Figure 2(b) shows the evolution of  $\phi_s(z,t)$ in short time after the swelling stars.  
It is seen that in the induction period where $\phi_s(z,t)$ is less than $\phi_{\rm cri}$, $\phi_s(z,t)$ evolves as in standard diffusion.   This can be understood as follows.
In the induction period,  $\xi$ and $\eta$ in eq.~\eqref{eqn:EvoEqns1} are equal to 
$\xi_g$ and $\eta_g$ and are constant. Furthermore, $\alpha(t)$ is close to the 
initial value $\phi_0/3=0$, and can be ignored. Therefore 
eq.~\eqref{eqn:EvoEqns1} reduces to the following linear equation for $\phi_s(z,t)$.
\begin{equation}
	\xi_g\dot \phi_s=  ( K + \frac{4}{3}G )\frac{\partial^2 \phi_s}{\partial z^2} 
	                      + \eta_g \frac{4}{3}\frac{\partial^2 \dot \phi_s }{\partial z^2} 
	                                                   \label{eqn:A1}
\end{equation}	
Equation \eqref{eqn:A1} can be solved by Fourier-Laplace transform under
the boundary condition \eqref{eqn:BoundaryConds}, which is now simplified 
as
\begin{equation}
	\frac{4}{3}\eta( \phi_s)  {\dot \phi_s } 
	  + K(\phi_s-\phi_{\rm eq}) + \frac{4}{3}G \phi_s = 0
	                                                \label{eqn:A2}
\end{equation}
where $\phi_{\rm eq}=3 \varepsilon_{\rm eq}$ is the equilibrium value of $\phi_s$ in the rubbery state.
Equation~\eqref{eqn:A2} is an ODE for $\phi_s(h_0,t)$, and can be solved easily. 
The solution of eq.~\eqref{eqn:A2} is shown in the inset of Figure 2(b). 
It is seen that $\phi_s(h_0,t)$ first increases slowly  with time, and then shoots up
to approach a steady state value  $\phi_{\rm st} =\phi_{\rm eq} K/(K + (4/3)G)$ 
when it reaches  the critical value $\phi_{\rm cri}$. The solution of eq.~\eqref{eqn:A1}  under such time varying boundary condition gives the profile of $\phi_s(z,t)$ in the induction period ($t<0.5\tau_{\rm dif}$) shown in Figure 2(b). 

Such reasoning indicates that the induction time $t_{\rm ind}$ is 
given by the time during which the solution of eq.~\eqref{eqn:A2} 
increases from $0$ to $\phi_{\rm cri}$, and is given by
\begin{equation}
	  t_{\rm ind} \approx \frac{4}{3}\frac{\eta_g}{K} \frac{\phi_{\rm cri}}{\phi_{\rm eq}}
	                                               \label{eqn:A3}
 \end{equation}	
Equation~(\ref{eqn:A3}) indicates that $t_{\rm ind}$ 
is essentially given by the rheological relaxation time of the glassy state 
$\tau_{\rm rhe}=\eta_g/G$ since $K$ and $G$ are  of the same order of magnitude in the 
glassy state.  Equation \eqref{eqn:A3} will be compared with numerical results later.

{\bf Case II diffusion period.} The phenomena occurring at the gel surface in the induction period
keep occurring at the front of the rubbery region, i.e., at the glassy side of the
glass/rubber boundary,  where the solvent supply from the rubbery side 
changes the glassy side to rubbery state. After the time of  $t_{\rm ind}$, the
region of length $h_w^{0} \simeq \sqrt{D_g t_{\rm ind}}$ in the glassy side
is changed to rubbery state,  where $D_g=(K+\frac{4}{3}G)/\xi_g$ is the 
diffusion constant in the glassy side, and the front moves the distance $h_w^{0}$.
Therefore the front velocity $v_{\rm front}$ is given by $h_w^{0}/ t_{\rm ind}$, or
\begin{equation}
     v_{\rm front} \simeq \sqrt{\frac{D_g}{t_{\rm ind}}} 
     	     \simeq \sqrt{\frac{ D_g K}{\eta_g} }
                                            \label{eqn:A4}
\end{equation}	
This scaling relation is consistent with the previous argument by Thomas and Windle model~\cite{QIAN2000CaseII,TW1982CaseII,HuiCY1987CaseII2}.

Figure 2(c) shows that at the end of the case II diffusion period, 
$v_{\rm front}$ starts to increase because the rubbery front coming from the two 
sides of the sample meets, and accelerate the permeation speed.  When the solvent
concentration at $z=0$ exceeds $\phi_{\rm cri}$, the glassy region quickly 
disappears.  The dynamics of this last stage will be discussed in the following. 

{\bf Relaxation period.} So far we have been discussing the motion in $z$ direction (the transverse direction). 
We now discuss the motion in $x$ and $y$ direction (the lateral direction).  
Figure 3 shows the result of the numerical calculation for the transverse strain
$\varepsilon_z=\Delta h(t)/h_0$ (black line) and the lateral strain 
$\varepsilon_x=\alpha(t)$ (red line), for two different 
parameter sets: Figure 3(a) is for  $\eta_r/\xi_r h^2_{0}=0.005$ and 
Figure 3(b) is for larger  viscosity, $\eta_r/\xi_r h^2_{0}=0.05$.   
It is seen that when the transverse strain $\Delta h(t)/h_0$ 
is increasing, the lateral strain $\alpha$ is also increasing.   
This is quite natural because when a gel is swelled, 
it expands in all directions. In the present situation, the lateral 
strain $\varepsilon_x$ is much smaller 
than the transverse strain $\varepsilon_z$ since the lateral expansion is constrained 
by the glassy region located at the center. In the middle of the case II diffusion period,
$\varepsilon_z$ becomes larger than the equilibrium value $\varepsilon_{\rm eq}$, but 
$\varepsilon_x$ remains much smaller than $\varepsilon_{\rm eq}$.  
When the two rubbery regions meet at the center,  $\varepsilon_z$ starts to
decrease and shows the undershoot, while $\varepsilon_x$ keeps increasing.  
Such behaviors were  reported in previous experiments~\cite{TW1981DiffusionMO}, but no 
theoretical analysis has been done.   Our theory is the first to explain the
behavior in a single theoretical framework. 

The time dependence of the thickness $\Delta h(t)$ is not simple:  In the case of  
Figure 3(a), $\Delta h(t)$ shows overshoot and undershoot before it relaxes to the final equilibrium value, 
while in the case of Figure 3(b), $\Delta h(t)$ does not show such undershoot.   
Such distinction can be qualitatively explained as follows. 

The relaxation of the volume strain $\varepsilon_z+2\varepsilon_x$ is governed by the 
diffusion of solvent, which  is characterized by the diffusion time $\tau_{\rm dif}$. On the
other hand, the relaxation of shear strain is governed by the viscoelasticity of polymer
in the rubbery state, which is characterized by the rheological relaxation time 
$\tau_{\rm rhe}= \eta_r/G$.  If $\eta_r$ is small,  the shear strain 
$\varepsilon_z - \alpha(t) $ quickly relaxes to zero, and the resulting isotropic strain
$\varepsilon_z  = \alpha(t) $ relaxes to the equilibrium value $\varepsilon_{\rm eq}$ 
with diffusion time  $\tau_{\rm dif}$. This gives the behavior shown in Fig.~3(a).  
On the other hand, if $\eta_r$ is large, $\varepsilon_z $  
and $\alpha(t) $ relax to their equilibrium values independently of each other with 
rheological relaxation time $\tau_{\rm rhe}$. This gives the behavior 
shown in Fig.~3(b).

%
\begin{figure}[tp]
	\begin{center}
	\includegraphics[width=0.46\textwidth,draft=false]{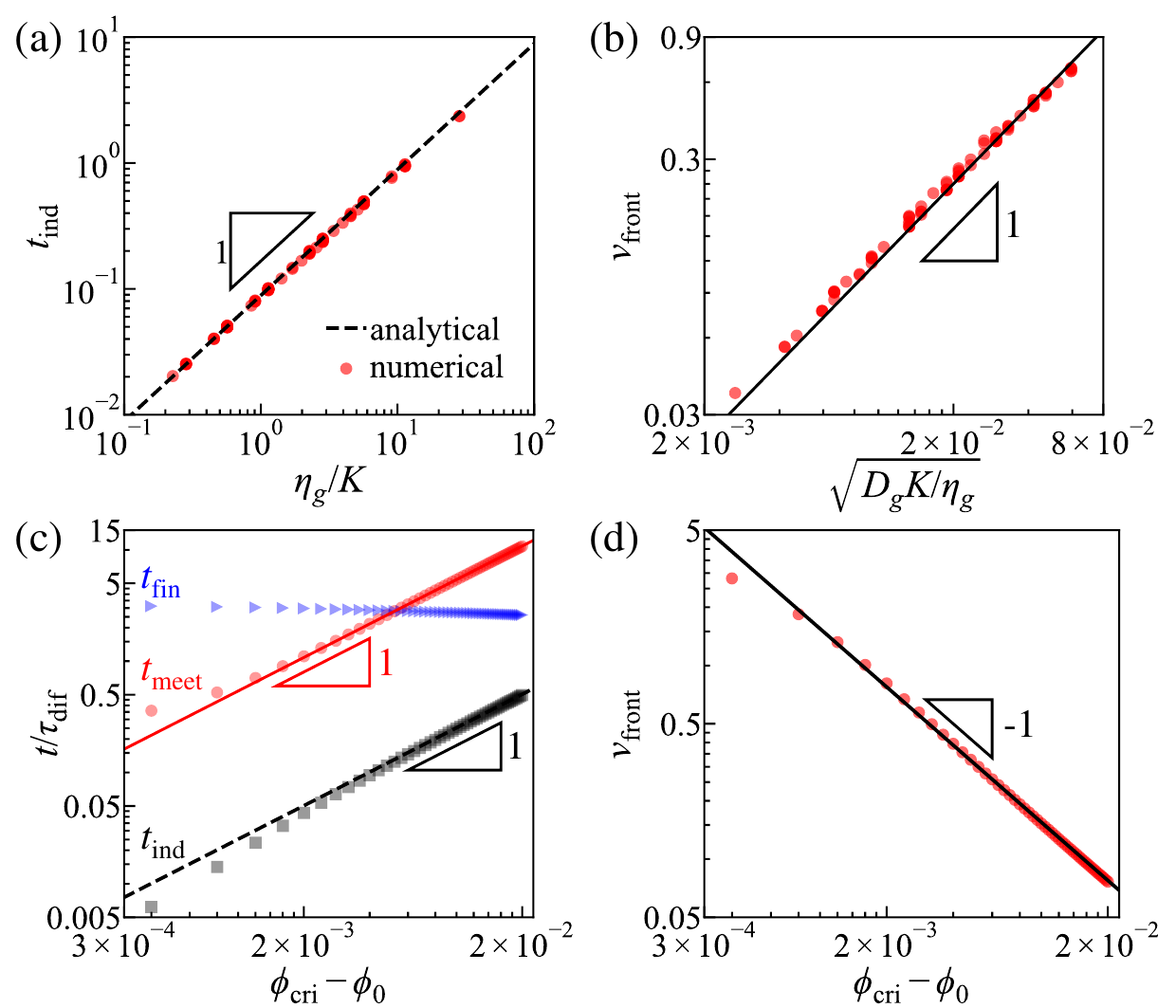}
	\caption{
		(a) The dependence of induction time $t_{\rm ind}$ on material parameters $\eta_g/K$ (in units of $\tau_{\rm dif}$). The black-dashed line is analytical solution of eq.~\eqref{eqn:A3} for $\phi_0=0$.  
		(b) The dependence of the front velocity $v_{\rm front}$ on $\sqrt{D_g K/\eta_g}$ (in units of $h_0/\tau_{\rm dif}$) for $\phi_0=0$. 
		(c) The dependence of induction time $t_{\rm ind}$ (black squares), case II diffusion time $t_{\rm meet}$ (red circle), and relaxation time $t_{\rm fin}$ (blue triangles) on $\phi_{\rm cri}-\phi_0$. The black-dashed line is obtained by $t_{\rm ind} =\frac{4}{3} \frac{\eta_g}{K} \frac{\phi_{\rm cri}-\phi_0}{\phi_{\rm eq}}$.
		(d) The dependence of the front velocity $v_{\rm front}$ (in units of $h_0/\tau_{\rm dif}$) (red circle) on $\phi_{\rm cri}-\phi_0$.
		 All data points are numerically calculated from the full model, while all solid lines are guiding lines.
	 }
	\label{fig4}
	\end{center}
\end{figure}
%

{\bf Material parameter dependence.} The case II diffusion is characterized by two physical quantities, the induction time
$t_{\rm ind}$ and the front velocity $v_{\rm front}$. We have shown that if the sample includes no solvent initially, i.e., if $\phi_0 $ is zero, $t_{\rm ind}$  and $v_{\rm front}$ 
are given by eq.~\eqref{eqn:A3} and  \eqref{eqn:A4}. Figure~4(a) and 4(b)  show the comparison of these equations with numerical results. 
They  confirm the validity of the scaling relations $t_{\rm ind} \sim \eta_g/K$ and 
$v_{\rm front} \sim \sqrt{ D_g K/\eta_g}$.

Figure 4(c) shows how the characteristic times of the three periods change when the initial solvent volume fraction $\phi_0$ is increased. It is seen that $t_{\rm ind}$ decreases with the increase of $\phi_0$ in proportion to $\phi_{\rm cri}-\phi_0$. Solving the boundary condition eq.~\eqref{eqn:BoundaryConds} under the initial condition $\phi_s=\phi_0$ 
and the approximation $\eta(\phi_s) \simeq \eta_g$ and $\alpha=\phi_0/3$, we have 
$t_{\rm ind} \approx \frac{4}{3}\frac{\eta_g}{K}\frac{\phi_{\rm cri}-\phi_0}{\phi_{\rm eq}}$.
Figure~4(d) indicates that $ v_{\rm front}$  increase with the increase of   $\phi_0$
satisfying the scaling relation $ v_{\rm front} \sim  (\phi_{\rm cri} - \phi_0)^{-1}$.  
This is not consistent with the argument to derive eq.~\eqref{eqn:A4}, which gives the scaling
relation $ v_{\rm front} \sim  (\phi_{\rm cri} - \phi_0)^{-1/2}$.  
This discrepancy arises from the inaccurate estimation for the precursor
length $h_w$: previously we assumed that $h_w$ is given by
$\sqrt{D_g t_{\rm ind}}$, which decreases with the increase of $\phi_0$.
On the other hand,  the numerical calculation indicates that
$h_w$ is almost independent of $\phi_0$, and is equal to $h_{w}^0$~\cite{SM}. 
If we use this result,  $v_{\rm front}$ is estimated by $h_w^0/t_{\rm ind}$ that is
\begin{equation}\label{eqn:vFront} 
	v_{\rm front} \sim \sqrt{\frac{D_g K}{\eta_g}} \frac{1}{\phi_{\rm cri}-\phi_0}
\end{equation}
As far as we know there are no experimental reports on how the front velocity changes
when the initial solvent volume fraction is changed.  Eq.~\eqref{eqn:vFront} is a 
prediction of our model, and it may be checked experimentally.

Figure~4(c) shows other characteristic times  defined in
Fig.~2(c): $t_{\rm meet}$ is the duration time of the case II diffusion stage,  and
$t_{\rm fin}$ is the relaxation time in the final relaxation stage.   
$t_{\rm meet}$ is estimated by $h_0/ v_{\rm front}$, 
and is therefore proportional to $\phi_{\rm cri} - \phi_0$. 
$t_{\rm fin}$ is estimated by $ t_{\rm fin}\sim h_0^2/D_r$ and is independent of $\phi_0$.  Such characteristic times depend on the sample thickness $h_0$:  
$t_{\rm ind}$ is independent of $h_0$, and $t_{\rm meet}$ and
$t_{\rm fin}$ are proportional to $h_0$ and $h_0^2$ respectively.   These 
relations are checked by the numerical calculation~\cite{SM}.


In summary, we proposed a continuum theory for the swelling of glassy polymer gels.
The theory describes the three dimensional coupling between diffusion and viscoelasticity,
and explains the classical experimental results found by Thomas 
and Windle.  The theory gives predictions  how the induction time and the 
front velocity depend on material parameters (diffusion constant and relaxation times) and sample condition (sample thickness and initial solvent concentration). These predictions can be checked experimentally. 

Since the theory can describe the glass transition induced by solvent diffusion and
mechanical forces, it may be used for a wider class of problems such as
as plastic deformation, membrane filtration and fracture or healing.

\section*{Associated Content}
{\bf Supporting Information}: detailed derivation of our theoretical model, and explanation on how we numerically obtain the front velocity and the characteristic times.

\section*{Acknowledgments}
We thank Hugues Chat\'{e} for useful discussions. This work was supported in part by the National Natural Science Foundation of China (NSFC) under grant No.~21961142020, the Fundamental Research Funds for the Central University under grant No.~YWF-22-K-101. We also acknowledge the support of the High-Performance Computing Center of Beihang University.


\newpage
\vskip 0.5truecm

\begin{figure*}[h]
  \centerline{for Table of Contents use only}
  \vskip 0.5truecm
  \centerline{\bf A Unified Model for Non-Fickian Diffusion and Anomalous Swelling of Glassy Polymer Gels}
  \vskip 0.5truecm
  \centerline{\it Peihan Lyu, Zhaoyu Ding, Masao Doi*, Xingkun Man*}
  \vskip 0.5truecm
	\begin{center}
		\includegraphics[scale=0.5,draft=false]{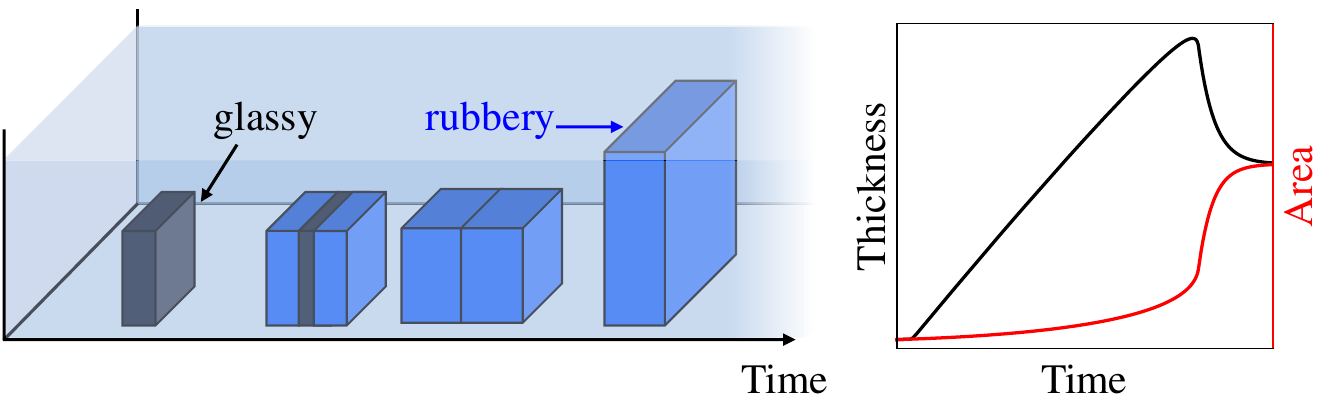}
	\end{center}
\end{figure*}

\end{document}